# On the bio-habitability of M-dwarf planets

A. Wandel

The Racach Institute of Physics, The Hebrew University of Jerusalem, 91904, Israel
amri@huji.ac.il

**Abstract**

The recent detection of Earth-sized planets in the habitable zone of Proxima Centauri, Trappist-1 and many other nearby M-type stars has led to speculations, whether liquid water and life actually exist on these planets. To a large extent, the answer depends on their yet unknown atmospheres, which may though be within observational reach in the near future by JWST, ELT and other planned telescopes. We consider the habitability of planets of M-type stars in the context of their atmospheric properties, heat transport and irradiation. Instead of the traditional definition of the habitable zone, we define the bio-habitable zone, where liquid water and complex organic molecules can survive on at least part of the planetary surface. The atmospheric impact on the temperature is quantified in terms of the heating factor (a combination of greenhouse heating, stellar irradiation, albedo etc.) and heat redistribution (horizontal energy transport). We investigate the bio-habitable domain (where planets can support surface liquid water and organics) in terms of these two factors. Our results suggest that planets orbiting M-type stars may have life-supporting temperatures, at least on part of their surface, for a wide range of atmospheric properties. We apply this analyses to Proxima b and the Trappist-1 system. Finally we discuss the implications to the search of biosignatures and demonstrate how they may be used to estimate the abundance of photosynthesis and biotic planets.

**Keywords:** astrobiology, M type stars; extra solar planets; habitable planets;

## 1. Introduction

Recent data of the Kepler mission show that 10-50% of the M-dwarfs have small planets orbiting within their habitable zone (Batalha *et al*. 2013; Dressing and Charbonneau, 2015), which implies that such planets could be found within less than 10 light years from Earth (Wandel, 2015; 2017a). A high abundance of Earth-size volatile-rich planets close to low mass M-stars is also supported by theoretical arguments and planetary formation models (Alibert and Benz 2017). Transiting planets of M-stars are also easier targets for the detection of biosignatures (Seager and Deming 2010). The detection of an Earth-sized planets within the habitable zone of Proxima Centauri (Anglada-Escude *et al.* 2016) and the Trappist-1 system (Gillon *et al.* 2017) has driven much interest in the question of whether planets orbiting red dwarf stars could support life, in spite of such planets being gravitationally locked with an eternal day and eternal night hemispheres (Heller, Leconte and Barnes 2011; Leconte, Wu, Menou and Murray 2015). Other threats to life come from the enhanced EUV radiation and X-ray flares of the host star (e.g. Ribas *et al.* 2016), greenhouse run-away and water loss (Kopparapu 2013; Luger and Barnes 2015). Over the last decade many authors suggested that these factors would be less severe than previously thought (Tarter *et al.* 2007; Scalo *et al.* 2007; Seager and Demming 2010; Gale and Wandel 2017). Although planets in the habitable zone of M-type stars are irradiated by much higher EUV and X-ray fluxes than Earth, many evolutionary models of M-dwarfs lead to water-rich planets. Furthermore, a prime atmosphere massive enough could survive the extended erosion during the energetic early evolution of M-dwarfs (e.g. Tian 2009).

A major condition to the evolution of life is the capability to support liquid water and surface temperatures in the range adequate for complex organic molecules. This condition depends not only on the irradiation from the host stat, but to a large extent on the effect of the planet's atmosphere. Global Climate Models using radiative transfer, turbulence, convection and volatile phase changes can calculate the conditions on planets, depending on the properties of their atmospheres. Such 3D climate models of M-dwarf planets show a presence of liquid water for a variety of atmospheric conditions (e.g. Pierrehumbert 2011a, Wordsworth 2015). Climate modeling studies have shown that an atmosphere only 10% of the mass of Earth's atmosphere can transport heat from the day side to the night side of tidally locked planets, enough to prevent atmospheric collapse by condensation (Joshi et al. 1997; Tarter et al. 2007; Scalo et al. 2007; Heng and Kopparla 2012). On locked planets the water may be trapped on the night side (e.g. Leconte 2013), but planets with enough water or geothermal heat, part of the water remains liquid (Yang et al. 2014). 3D GCM simulations of planets in the habitable zone of M-dwarfs support scenarios with surface water and moderate temperatures (Yang et al. 2014; Leconte et al. 2015; Owen and Mohanty 2016; Turbet *et al.* 2016; Kopparapu et al. 2016; Wolf 2017 to name a few)**.** A variety of surface temperature distributions emerges. While rocky planets with no or little atmosphere, like Mercury, have an extremely high day-night contrast and planets with a thick, Venus-like atmosphere tend to be nearly Isothermal, intermediate cases, with up to 10 bar atmospheres conserve significant surface temperature gradients (e.g. Selsis et al. 2011). Wolf et al. (2017) find (for F-K type hosts) that stable habitable ocean worlds can be maintained for global mean surface temperatures in the range 235-355 K.

It is instructive to consider whether similar results can be obtained more generally, using basic principles. Lower order climate models for synchronously rotating planets have been discussed by several authors (Haberle et al. 1996; Pierrehumbert 2011a; Heng and Kopparla 2012, Mills and Abbot 2013; Wordsworth 2015; Koll and Abbot 2016).

Rater than the usual treatment of the habitable zone, defined as the region around a star where planets can support liquid water (e.g. Kasting, Whitmire and Reynolds 1993), we investigate the bio-habitable zone, where surface temperatures can support, in addition to liquid water, also complex organic molecules (Wandel and Gale 2017).

As the properties that affect the atmosphere and surface temperature distribution are complex, entangled and difficult for observational determination and delineation, we combine them in merely two factors: redistribution and atmospheric heating. In this two-dimensional parameter space we chart the domain where liquid water and life supporting temperatures are present on at least part the planet's surface.

In section 2 we derive a simple model for the surface temperature distribution of tidally locked planets in terms of the radiative and atmospheric heating and the horizontal heat transport. In section 3 we define the heating factor of the planet's atmosphere. Section 4 demonstrates the model parameters for the terrestrial planets in the Solar System and section 5 applies the model to the Trappist-1 planets. In sections 6 and 7 we define the bio-habitable zone and investigate the its dependence on the atmospheric properties in terms of the heating factor and the heat transport. Finally, in sections 8 and 9 we consider the implications to biosignature detection and estimate the abundance of biotic planets.

## 2. Thermal models for tidally locked planets

Planets within the habitable zone of M-dwarfs are often close enough to their hosts to be tidally locked. In synchronously orbiting planets, horizontal energy transport has a major impact on the surface temperature, as it determines the horizontal temperature gradient and the temperature on the night hemisphere. Overall the surface temperature is determined by three major factors: irradiation from the host star (insolation), atmosphere (transmission and greenhouse effect) and horizontal heat transport. Energy redistribution due to convection can in principle be determined from the atmospheric properties (atmospheric pressure, heat capacity and wind speed, global circulation patterns), which may be identified with broadband thermal phase curves, e.g. Wang et al. 2014. However, these data may be difficult to obtain and disentangle in exoplanets. Hence we describe the heat redistribution in a parametric form, with a global redistribution parameter ($f$) and a local one ($b$). We subsequently show that these two parameters are related.

Following previous works (Haberle et al 1996; Goldblatt 2016), we separate the energy balance into day and night hemispheres, with advective interchange of energy.
Atmospheric temperatures of aquatic locked planets may be nearly uniform horizontally, as approximated by the Weak Temperature Gradient model (Joshi et al. 1997; Merlis & Schneider 2010), while dry, low surface pressure, tidally locked planets may have larger horizontal temperature differences, in particular between day and night side (Koll and Abbot 2016). Intermediate cases can be described by a global energy redistribution scheme (e.g. Pierrehumbert 2011a). In addition to global redistribution we consider local heat transport, representing small scale advection and turbulence in the atmospheric surface layer, which has been demonstrated by 3D models to be essential for the global energy balance on tidally locked planets (Wordsworth 2015).

### 2a. Tidally locked planets with global redistribution

Without atmosphere the surface temperature of a slowly rotating planet is determined only by the irradiation from the host star and the latitude, that is, the angular distance $\theta$ from the substellar point,

$$\sigma T^4(\theta) = (1 - A)SF(\theta). \qquad (1a)$$

Here $\sigma$ is the Stefan Boltzman constant, $A$ is the Bond albedo and $F=\cos(\theta)$. The irradiation or insolation is given by $S=L_*/4\pi a^2$, where $a$ is the distance of the planet from its host star and $L_*$ is host star's luminosity[1].

Planets with an efficient horizontal heat spread or rapidly rotating ones are nearly isothermal, the equilibrium surface temperature being

$$T_{eq} = [(1 - A)S/4\sigma]^{1/4}. \qquad (1b)$$

The atmosphere has two main effects on the surface temperature: vertical energy exchange between the atmosphere and the surface, and horizontal heat transport. The former can be described by an energy balance equation of the form (e.g. Mills and Abbot 2013)

$$(1 - A)SF(\theta) + e_a\sigma T_a^4 = \sigma T^4 + c_a(T - T_a),$$

where $T_a$ is the atmospheric temperature, $e_a$ is the atmospheric long-wave emissivity and $c_a$ is the surface-to-atmosphere exchange constant. In the following we replace the two unknown parameters $e_a$ and $c_a$ by a single one (the heating factor $H$, sec. 2b), in which we include also the insolation and the albedo. We reduce the extra variable $T_a$ by assuming the atmosphere is horizontally isothermal (the Weak Temperature Gradient model, which has been shown to be a very good approximation in slowly rotating planets, Pierreumbert 2011a,b), and by implicitly replacing the last term on the RHS with the horizontal redistribution mechanism[2].

Horizontal energy redistribution transports heat across the temperature gradient created by latitude-dependent irradiation. We describe this effect by assuming that at every latitude of the day side a fixed fraction $f$ of the heat due to the radiation from the host star is removed and homogeneously distributed over the entire planet,

$$F(\theta) = \begin{cases} \dfrac{f}{4} + (1-f)\cos(\theta) & 0 \leq \theta \leq 90° \\ \dfrac{f}{4} & 90° < \theta \leq 180° \end{cases} \qquad (1c)$$

---

[1] Alternatively, $S$ may be expressed in the form $S=\sigma T_*^4 R_*^2/a^2$, where $T_*$ and $R_*$ are the host star's surface temperature and radius, respectively. The temperature can then be expressed in the form $T_{eq} = T_*(R_*/2a)^{1/2}(1 - A)^{1/4}$.

[2] The combined effect of a vertical energy exchange with the atmosphere and an atmospheric horizontal circulation can be shown to have the approximate effect of horizontal heat redistribution, e.g. Koll and Abbot 2016.

where $\theta$ is the latitude, measured from the sub-stellar point (there $\theta=0$, while $\theta=90°$ is the terminator, separating the day and night hemispheres). The global redistribution parameter may be in the range $0 \leq f \leq 1$ and is primarily determined by the atmospheric pressure and circulation. 3D climate models show that $f$ is influenced also by the planet's rotation, size and surface friction (e.g. Carone et al 2015).

Geothermal heating of the order of that of Earth, 0.1 W/m$^2$(Davies & Davies 2010), may be neglected, as it is less than $10^{-3}$ of Earth's irradiation. However, for planets in the inner HZ of M-stars tidal heating may be important (Dobbs, Heller and Turner 2017). This issue will be considered in a subsequent work.

Solutions of eq. 1a are shown in fig. 1. We have used the insolation of Proxima Centauri b (Anglada-Escude *et al*. 2016), $S = 0.65 S_\oplus$, where $S_\oplus = 1362$ Watt/m$^2$ is the solar constant at the Earth's orbit. We have assumed an albedo of $A=0.1$ (like the Moon). Cloud (or ice) coverage (e.g. Yang et al. 2014) would raise the albedo, lowering the temperature. For example, if $A=0.3$ (like Earth) the curves in fig. 1 need to be multiplied by 0.94.

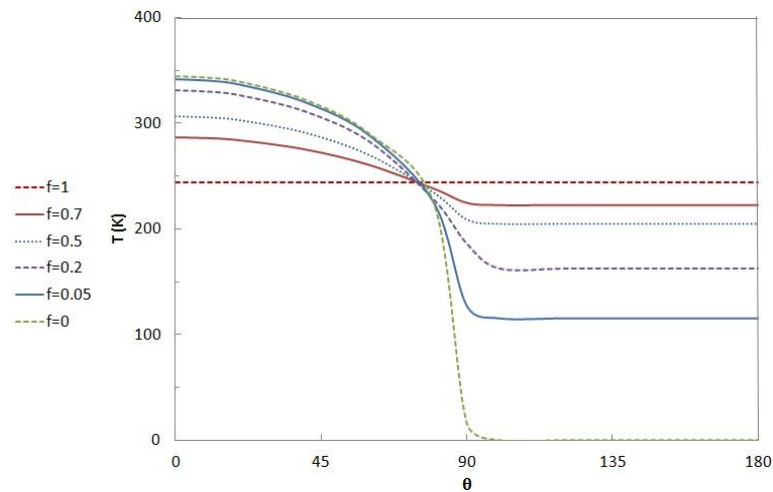

Fig. 1. Surface temperature distribution for various global heat redistribution rates, with the insolation of Proxima b. Equilibrium temperature is plotted vs. latitude for different values of the global redistribution factor $f$. When $f=1$ (full global redistribution) the planet becomes isothermal. A small amount of local heat redistribution has been assumed in order to avoid unrealistically sharp temperature changes.

**2b. The atmospheric impact**

The atmosphere can have a major impact on the surface temperature. We follow the analyses of Guillot (2010), adapting it to rocky planets. Part of the stellar irradiation is absorbed and scattered by the atmosphere, which may be described by the parameter $\alpha$, defined as the fraction of short waveband irradiation at the top of the atmosphere reaching the surface.

Strictly speaking, $\alpha = \exp[-\tau_{SW}/\cos(\theta)]$, where $\tau_{SW}$ it is the vertical optical depth of the atmosphere in the short waveband. However, assuming the atmosphere is relatively transparent in the short waveband, this expression may simplified by taking a weighted average over latitude. Values of $\alpha$ for the terrestrial planets of the Solar System are given in table 1.

The atmospheric attenuation of the planet's radiative cooling, mainly in the infrared, is determined by the abundance (partial pressure) of greenhouse gases, such as $CO_2$, $CH_4$, $N_2O$ and $H_2O$. We define the greenhouse heating parameter $H_g$ as the amount of increase in the thermal energy at the surface due to the greenhouse effect. It can be shown (e.g. Guillot 2010) that the greenhouse heating is related to the optical depth,

$$H_g = 1 + \tau_{LW}$$

where $\tau_{LW}$ is the optical depth of the atmosphere in the long wavelength regime. A planet with no atmosphere, or with no greenhouse gases has $H_g=1$. For an optically thick atmosphere, $f \approx 1$ and the planet becomes nearly isothermal, with

$$T_{isoth} = T_{eq} (\alpha H_g)^{1/4}, \qquad (2a)$$

and $T_{eq}$ is the equilibrium isothermal surface temperature (eq. 1b). This result resembles the scaling relation derived by Pierrehumbert (2011b) for the surface temperature of an isothermal planet with an optically thick atmosphere[3]. The lowest temperature on the night side (at the point opposite to the substellar one, $\theta=180°$) eqs. 1b and 1c give

$$T_{min} = 278 \, [f(1-A) H_g \alpha S/S_\oplus]^{1/4} \text{ K}. \qquad (2b)$$

Since in the long wave band radiation is scattered in all directions, the greenhouse effect redistributes heat not only vertically but also horizontally. For an optically thin atmosphere, $(\tau_{LW} <1)$, $\alpha H_g \sim 1$ and the effective redistribution is approximately $f \sim \tau_{LW}$, hence eq. 2b gives

$$T_{min} \sim T_{eq} \tau_{LW}^{1/4}.$$

A similar lower limit, $T_{min}=T_{eq}(\tau_{LW}/2)^{1/4}$, has been found by Wordsworth (2015) for the nightside temperature of locked planets with optically thin atmospheres.

**2c. Local heat redistribution**

In addition to governing global heat redistribution, atmospheric advection can transport heat locally by small scale advection. Altogether we add to eq. 1a a term of local heat transport (LHT) denoted by $\dot{Q}$, in addition to the factors of atmospheric screening ($\alpha$) and greenhouse heating ($H_g$) discussed above, giving

---

[3] For an optically thick atmosphere $H_g \sim \tau_{LW}$, hence eq. 2a gives $T \sim T_{eq} \tau_{LW}^{1/4}$. Pierrehumbert's relation is $T \sim T_{eq} \tau_{LW}^{\beta}$, where $\beta$ depends mainly on the molecular weight. Typical values for $N_2$ or $CO_2$ atmospheres are $\beta \sim 0.1-0.3$.

$$\sigma T^4 = (1-A)H_g \alpha S F(\theta) + \dot{Q}. \qquad (3)$$

LHT transports energy in the direction perpendicular to the temperature gradient, hence the heat flux at a given latitude is proportional to the negative derivative of the temperature. The overall energy flow is proportional to $-R\sin(\theta)\frac{dT}{dx}$, or $-\sin(\theta)\frac{dT}{d\theta}$, where $x$ is the horizontal coordinate, $R$ is the planet's radius and the sine term gives the dependence of the surface area element on the latitude. The net heating is the derivative of the energy flow,

$$\dot{Q} = -B\sin(\theta)\frac{d^2 T}{d\theta^2}$$

where $B$ is the advection parameter (cf. Haberle et al. 1996; Goldblatt 2016)

$$B = \frac{p_s c_p u}{gR}. \qquad (4)$$

Here $p_s$ is the surface atmospheric pressure, $c_p$ – the heat capacity (e.g. for air $c_p$=1 kJ Kg$^{-1}$ K$^{-1}$), $u$ is the wind speed and $g$ – the surface gravity. For example, on Earth at sea level $B$= 15 ($u$/10ms$^{-1}$) Wm$^{-2}$K$^{-1}$.

Substituting these expressions for $\dot{Q}$, eq. 3 gives

$$\sigma T^4 = (1-A)H_g \alpha S F(\theta) - B\sin(\theta)\frac{d^2 T}{d\theta^2}. \qquad (5)$$

Solving eq. (5) for the temperature at $\theta$=0 (the substellar point, using eq. 1c) gives

$$T(0) \equiv T_0 = \left[\sigma^{-1}(1-A)H_g \alpha S(1-\tfrac{3}{4}f)\right]^{\frac{1}{4}}, \qquad (6a)$$

while at the opposite point (eq. 2b)

$$T(\theta\text{=}180°) = \left[(1-A)H_g \alpha S f/4\sigma\right]^{\frac{1}{4}}. \qquad (6b)$$

**2d. The dimensionless energy balance equation**

Eq. 5 can be written in a dimensionless form. Defining a normalized temperature $y=T/T_0$ and dividing by $(1-A)H_g \alpha S$ gives

$$(1-\tfrac{3}{4}f)y^4 = F(\theta) - b\sin(\theta)\frac{d^2 y}{d\theta^2}, \qquad (7)$$

where the dimensionless local heat transport parameter $b$ is given by

$$b = \frac{B T_0}{(1-A)H_g \alpha S}. \qquad (8)$$

Eq. 7 can be solved numerically (figs 2,3) with the boundary condition $y(0)=1$, $y'(0)=0$. Physically *b* is the advection parameter multiplied by the ratio of temperature to radiative heating at the substellar point. This choice of normalization is supported by 3D simulations which suggest that the local advection increases with radiative heating (Wordsworth 2015). As an example, the Earth atmosphere at sea level has $b \approx 6$. Typical values of *b* for the terrestrial planets of the Solar System are listed in table 1.

Unlike[4] *f*, *b* is not a free parameter, as the local advective heat transport is determined by the planetary and atmospheric properties (eq. 4). To some extent, global and local heat redistribution are related; when the LHT is not small (*b>1*), heat is globally redistributed (cf. fig. 3), hence large LHT may be approximated by increasing the global heat redistribution parameter (cf. . The two processes are complemetary, as global redistribution may have sources other than advection, while LHT smoothens temperature gradients on smaller scales. Although in some cases the two parameters, *f* and *b* can be combined, for the sake of generality we retain both heat redistribution terms in eqs. 5 and 7.

When *b<<1* (local heat transport is small compared to radiative heating), $\dot{Q}$ can be neglected and eq. 3 can be solved analytically,

$$T(\theta) = 394 \left( \frac{(1-A)H_g \alpha L}{a^2 L_\odot} F(\theta) \right)^{1/4} \text{K}, \qquad (9)$$

where *a* is the semi major axis of the planet's orbit expressed in AU and *L* is the host's luminosity. For a nearly circular orbit the insolation relative to Earth is given by $s = S/S_\oplus = (L/L_\odot)a^{-2}$. Eqs. 2a and 9 give

$$T(\theta)=1.42\ T_{eq}\ [\ HF(\theta)\ ]^{1/4},$$

where *H* is defined by eq 12. At the sub-stellar point $(\theta = 0°)$ eq. 9 coincides with eq. 6a. When the LHT is small (*b<1*), its impact on the global energy balance is small and the night side is nearly isothermal. This can be seen in figs 1 and 2. Far from the terminator the night side temperature approaches $T=(f/4)^{1/4}T_0$ hence $y(180°)=(f/4)^{1/4}$. Similar results are obtained by GCM calculations (e.g. Wordsworth et al. 2015; Koll and Abbot 2016).

---

[4] Obviously also global heat transport is determined by the atmospheric and planetary properties, as may be calculated by GCM models, but the relation is less explicit, depending i.a. on the vertical transport and on global circulation currents.

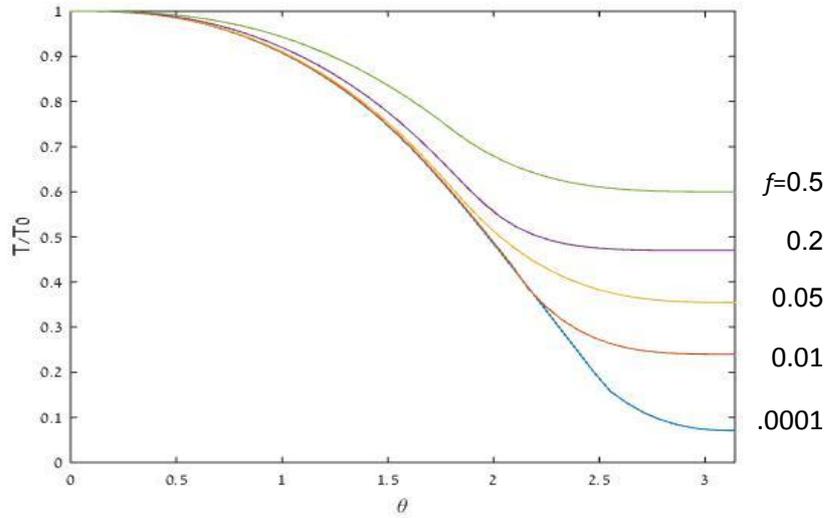

Fig. 2. Normalized temperature profiles for different values of the global redistribution parameter (marked on the right axis, $f=10^{-4}$, 0.01, 0.05, 0.2, 0.5) with moderate LHT ($b=0.5$). For large global redistribution ($f>0.3$) the normalization factor $T_0$ depends on $f$.

Numerical solutions of eq. (7) are shown in figs. 2 and 3. Note that in general $T_0$ depends on $f$ (eq. 6a). While for $f<0.3$ this dependence is negligible, for larger values of $f$ global heat redistribution can significantly reduce the substellar temperature.

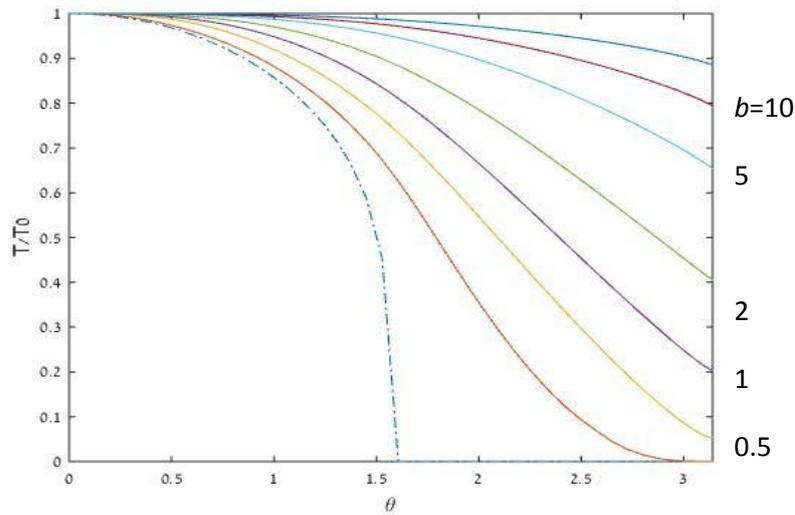

Fig. 3. Normalized temperature profiles for several values of the local transport coefficient ($b=0.1, 0.5, 1, 2, 5, 10$ and 20; some are marked on the right axis), with $f=0$. The dash-dotted curve corresponds to eq. 1a with zero redistribution. Note that for $b>1$ the normalization factor $T_0$ depends on $b$.

Fig. 3 shows that when the LHT is comparable or larger than the radiative heating from the host star ($b>1$), its impact on the energy balance cannot be neglected, as it transports significant amounts of energy from the day hemisphere to the night one. Energy conservation implies

$$2\pi R^2 \sigma \int_0^\pi T^4(\theta)\sin(\theta)d\theta = \pi R^2(1-A)H_g\alpha S. \qquad (10)$$

When *b<<1*, $T_0$ is given by eq. 6a and eq. 10 is satisfied. In general eq. 10 gives

$$T_0 = T_1 \left[\int_0^\pi y^4(\theta)\sin(\theta)d\theta\right]^{-1/4} \qquad (11)$$

where

$$T_1 = (HS_\oplus/2\sigma)^{1/4} = 331 \; H^{1/4} \; \text{K}$$

and *H* is the heating factor (eq. 12).

For large LHT (b>>1) the planet becomes nearly isothermal ( *y=1*), as in the case of *f=1* (eq. 2a). Also in this case it is easy to see that eqs. 10 and 11 are satisfied and we may replace *b* with an effective global heat redistribution *f*, as discussed after eq. 8.

**3. The heating factor**

The albedo and atmospheric transmissivity can be combined to give an effective reflectivity *(1-A)α*. In eqs. 8 and 9 this term appears together with the greenhouse factor and the insolation. Hence it is convenient to combine the albedo, the atmospheric screening and heating effects and the insolation, into a single dimensionless parameter (already used above), which we call the heating factor

$$H = (1-A)H_g\alpha s. \qquad (12)$$

Here $s=S/S_\oplus$ is the insolation relative to Earth and $H_g$ is the greenhouse factor. Fig. 4 shows surface temperature profiles of for various values of the heating factor.

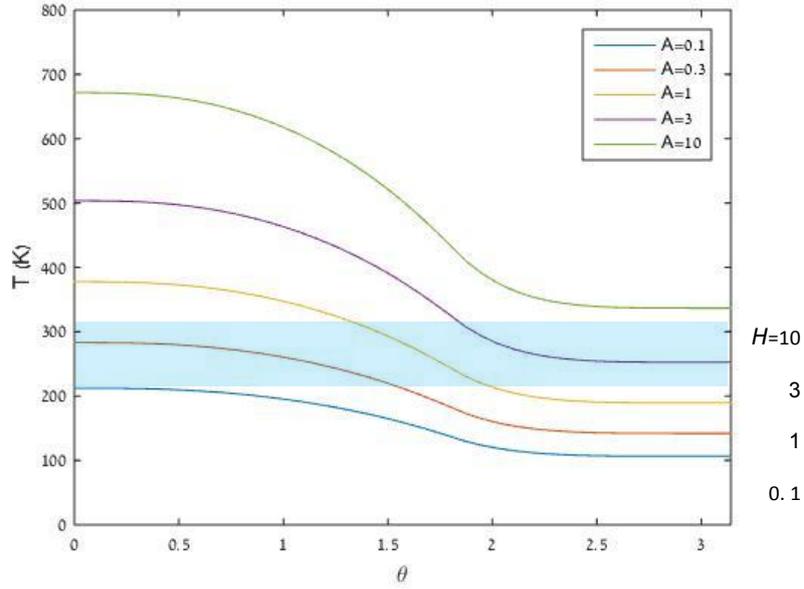

Fig 4. Temperature profiles for several values of the heating factor (from below, *H*=0.1, 0.3, 1, 3, 10). A global heat transport of *f=0.2* and a moderate *(b=0.5)* LHT are assumed. The liquid water temperature range at 1 bar is indicated by the shaded blue area.

As the planet's insolation is usually known, it is useful to leave it out of the total heating, defining the *atmospheric* heating factor, which describes only the atmospheric impact (and the albedo),

$$H_{atm} = (1 - A)H_g\alpha \ . \qquad (13)$$

Fig. 4, as well as figs. 6-7 below, can be modified to use the atmospheric heating factor rather than *H*, scaling the temperatures by $s^{1/4}$, namely $T(H_{atm}) = s^{-1/4}T(H)$. E.g. for Proxima b this factor is $(0.65)^{-1/4}=1.1$. This adjustment is applied to the Trappist-1 planets in fig. 6.

Atmospheric heating can have a strong influence on the existence of liquid water, but other processes may also play a role. Under certain conditions, water on tidally locked planets may be transported from the day side and deposited by precipitation at the night side, where it is trapped as ice (Leconte 2013; Menou 2013, Yang et al. 2014). However, as noted by these authors, complete trapping will occur only for planets with low water content and geothermal heat flux lower than that of Earth.

**4. The Solar System**

We demonstrate some values of the heating factors and the other parameters defined above for different atmospheres and insolations, by calculating them for the terresrial planets of the Solar System, for which the albedo and atmospheric data are well known. The atmospheric heating factor can be determined by comparing the effective surface temperature to the equilibrium temperature (the planet's surface temperature calculated without the atmospheric effects and A=0, that is, $H_{atm}$=1. For $L=L_O$ the equilibrium temperature is given by $T_{eq}$=278 $a^{-2}$K, where $a$ is the average distance from the Sun in AU.

|  | Earth | Venus | Mars | Mercury |
|---|---|---|---|---|
| *a* (AU) | 1 | 0.723 | 1.52 | 0.387 |
| $T_{eq}$(K) | 278 | 327 | 225 | 447 |
| *T*(K) | 288 | 737 | 210 | 100-700 |
| *A* | 0.28 | 0.9 | 0.25 | 0.068 |
| $\alpha$ | 0.48 | 0.1 | 1 | 1 |
| $\tau_{LW}$ | 2.3 | 2500 | 0.003 | 0 |
| *b* | 6 | 50 | 0.3 | 0 |
| *f* | 0.95 | 1 | 0.85 | 0.003 |
| *H* | 1.15 | 50 | 0.33 | 6.2 |
| $H_{atm}$ | 1.15 | 26 | 0.75 | 0.93 |

Table 1. Data and calculated atmospheric parameters for the terrestrial planets of the Solar System. $T_{eq}$ is the equilibrium surface temperature without atmosphere, *T* is the measured surface temperature and *A* is the Bond albedo. The greenhouse heating factor and longwave oprical depth $\tau_{LW}$ are derived from the measured albedo and transmissivity. Then the heating factors are given by $H_{atm}= (T/T_{eq})^4$ and $H= H_{atm}/a^2$.

Some indicative values of the albedo for common planetary surfaces are *A=0.15* (bare rock), 0.3 (patch water clouds, as in present Earth) and 0.6 (complete water cloud coverage). The greenhouse factor is calculated using the relation

$$H_g = H_{atm}/[(1-A)\alpha],$$

where the atmospheric heating factor is given by $H_{atm} = (T/T_{eq})^4$. For example, Earth's atmosphere (0.78 bar $N_2$, 0.21 $O_2$ and 375 ppm $CO_2$) has $H_{atm}$=1.15 and its short wave transmissivity is $\alpha$=0.48, hence $H_g = 3.3$ and $\tau_{LW}$ =2.3.

The larger the partial atmospheric pressure and the abundance of greenhouse gases, the stronger is the greenhouse effect. E.g. for Venus we get $\alpha H_g$= 250. In order to estimate the greenhouse factor of the Venerian atmosphere (~90 bar $CO_2$) we assume $\alpha$=0.1. For the airless Mercury we assume $\alpha=H_g$=1 and *f* may be determined from eq. 15a below.

The relatively fast rotating terrestrial planets (Earth and Mars) are nearly isothermal, so they may be considered to have an *effective* global redistribution parameter near unity (actually slightly less, because of the day-night differences). Also Venus is nearly isothermal, in spite of its nearly synchronous rotation, because of its dense atmosphere and cloud cover which efficiently smoothen surface temperature gradients. For Mercury *an effective* global redistribution parameter is estimated from the lowest surface temperature on the night side and eq. 2b. The parameter *b* is calculated from eq. 8 with *u* assumed to be *5* m/s for Earth, 20 m/s for Mars and *1* m/s for Venus.

For exoplanets the amount of heat redistribution and the albedo may be estimated by observing the thermal phase variation, which could be measured by JWST (Kreidberg and Loeb 2016). The albedo may also be determined by high contrast spectroscopy, which can be accomplished by just one night of the planned 40-meter ELT (Snellen *et al.* 2015).

Figure 5 shows the four terrestrial planets of the Solar System in the *T vs. H* diagram. Note that Venus, Earth and Mars have average temperatures consistent with the equilibrium (isothermal surface) green curve, while Mercury's temperature gradient is larger than that of a locked planet with global redistribution parameter of *f=0.1* (the dashed curves), consistent with the effective value of *f* calculated above.

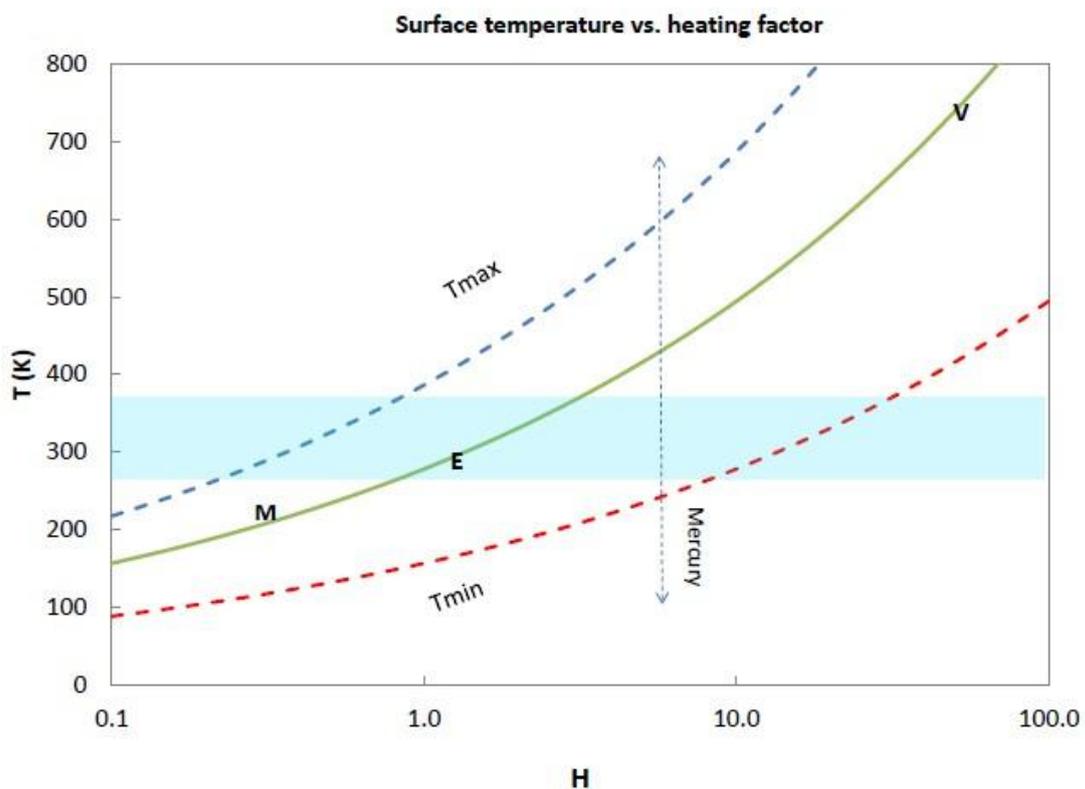

Fig. 5. The locations of the four terrestrial planets of the Solar System in the T vs. H diagram. The green solid curve shows the equilibrium temperature for an isothermal planet, while the dashed curves show the highest and lowest surface temperatures of a tidally locked planet with 10% global redistribution (*f=0.1*).

## 5. The Trappist-1 system

We demonstrate the derivation of the parameter values for exoplanets with a given irradiation but an unknown atmosphere for the Trappist-1 planets, calculating the bio-habitable range[5] of their heating factor. The insolations of the Trappist-1 planets can be determined from their measured periods (determining the distances from the host star). The insolation is calculated using the bolometric luminosity. As a caveat note that if $\tau_{LW}$ of the planet's atmosphere is not small, only a fraction of the irradiative flux penetrates the atmosphere and reaches the surface, as most of the luminosity Trappist 1a is in the long waveband range. Habitability and climate calculation of the Trappist-1 planets have been attempted by several authors using GCM-codes. For global ocean planets Wolf (2017) concludes that only Trappist-1e is likely to be habitable. However our model takes in account also local habitability on semi-dry planets, e.g. in water-belt worlds.

Fig. 6 shows the surface temperature distribution for each of the six inner planets of the Trappist-1 system, assuming a 10% global heat redistribution (*f=0.1*) and an atmospheric heating factor $H_{atm}=1$, similar to that of Earth's atmosphere. We note that the five inner planets could support liquid water and organic molecules on some part of their surface. Trappist-1g could be marginal, being almost entirely frozen, with an eyeball configuration of liquid water at the substellar point. With an atmosphere of a heating factor 4 times lower than that of Earth ($H_{atm}=0.3$, shown in fig. 6 by the lower dashed curve for Trappist-1g*)* all planets except f and g could support a bio-habitable region. For an atmospheric heating factor ~3 times larger than that of Earth ($H_{atm}=3$, upper dashed curve for Trappist-1b) without moist greenhouse runaway, the five outer planets could support a bio-habitable region, with Trappist-1b being marginal, having a night side temperature of nearly 373K.

An increased $H_{atm}$ may be due to a larger greenhouse factor (i.e. larger $\tau_{LW}$). As noted above, scattering of the incoming long waveband radiation from the host may reduce the amount of stellar radiation actually arriving at the planet surface, lowering the atmospheric screening parameter $\alpha$ and producing a negative feedback on the heating factor. For example, if the atmospheric scattering in the long waveband reduces the effective heating factor a by a factor of 3, we may conclude from fig. 6 that atmospheres with $H_{atm}$ as high as ~10 could support bio-habitable conditions for all six Trappist-1 planets.

---

[5] See section 6 for the definition of bio-habitability

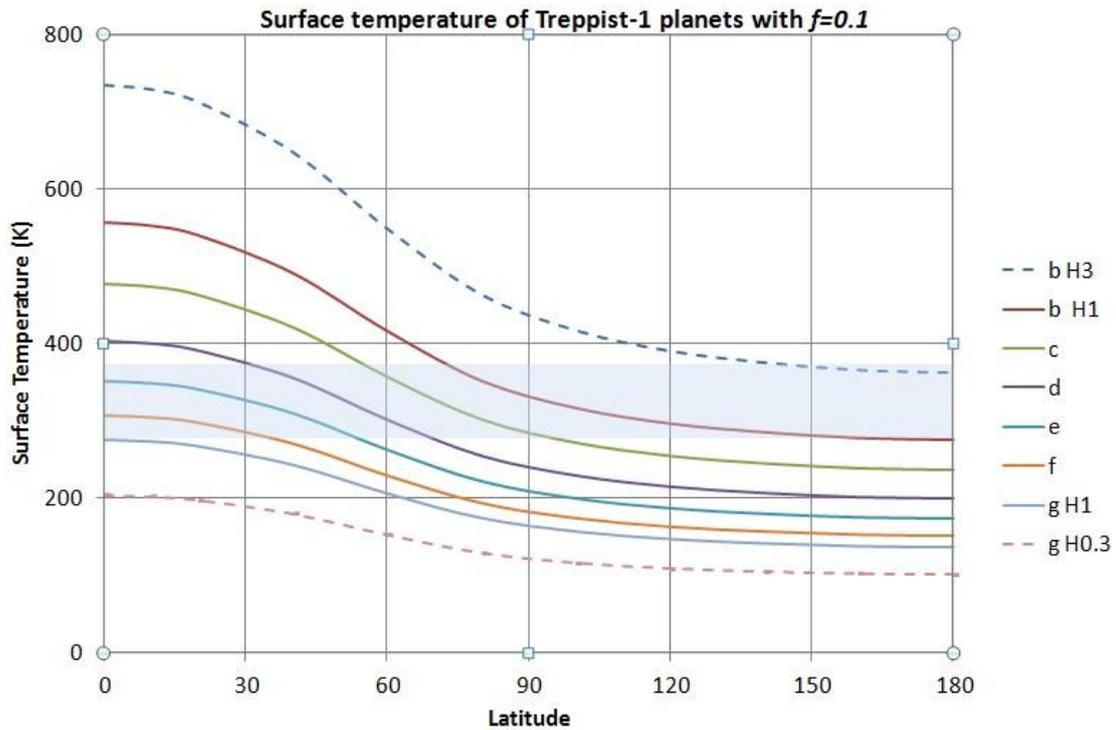

Fig 6. Surface temperature profiles for the six inner Treppist-1 planets. The global heat transport parameter is taken as $f=0.1$ and the LHT parameter - $b=0.5$. The continuous curves denote an atmospheric heating factor of $H_{atm}=1$. Dashed curves mark the coldest (g) and hottest (b) planets with $H_{atm}=0.3$ and $H_{atm}=3$, respectively. The temperature range of liquid water at 1 bar is indicated by the shaded blue area.

## 6. The bio-habitable zone

The traditional habitable zone is associated with the existence of liquid water on the planetary surface. The extent of the liquid water zone around the host star may be described by the Kombayashi-Ingersoll radiation limit (Kombayashi 1967, Ingersoll 1969, Kasting et al. 1993) also called the "Runaway greenhouse" limit. The inner edge of the habitable zone is located at the distance from the host star, where surface water is completely vaporized or at which water reaches the upper atmosphere, where, it can be dissociated by ultraviolet radiation (unless, as in Earth's atmosphere, there is cold-trapping of water at a tropopause-like layer). Greenhouse runaway processes may push this inner edge outwards, while cloud coverage may push it inwards to as small radii as about 0.5 AU for a Sun-like host (e.g. Kasting et al. 1993; Yang et al. 2016). The outer edge of the habitable zone is determined by water being completely frozen on the planet surface. For planets with a present Earth atmosphere and Sun-like host this limit may be not much larger than Earth's orbit, because of the positive feedback of runaway snowball effect. The outer limit may however be increased considerably, beyond 2 AU for a Sun-like host, by increasing the $CO_2$ atmospheric abundance, e.g. by out-gassing and geological $CO_2$ enrichment of the atmosphere (Forget 2013). The freezing point of water may also be lowered by high salinity.

In the following we will define the habitable zone not in the spatial sense, but rather in the climatic sense, quantified in the parameter space spanned by the heating factor and heat redistribution. Wolf et al. (2017) find four stable climate states defined by their global mean surface temperatures ($T_s$); snowball ($T_s \leq 235$ K), water belt (235 K $\leq T_s \leq$ 250 K), temperate (275 K $\leq T_s \leq$ 315 K), and moist greenhouse ($T_s \geq 330$ K), the three latter being able to maintain habitable ocean worlds. It has been claimed that the outer boundary of the HZ may be reduced by snowball limit-cycles, but Haqq-Misra et al. (2016) find that those cannot occur on K- or M-star planets.

We define the "Bio-habitable zone", which differs from the usual habitable zone in two ways. First, rather than quantifying the spatial boundaries where planets can globally support liquid water, we consider the domain in the parameter space, defined by atmospheric heating and heat redistribution, where liquid water and life supporting temperatures occur at least on part of the surface of a tidally locked planet.

Depending on the pressure, water may be liquid at temperatures far beyond the temperatures allowing organic complex molecules, the Komabayashi-Ingersoll limit may not be relevant for bio-habitability, in particular if oxygenic photosynthesis is considered (cf. Gale and Wandel 2017). We adopt 373K as a conservative upper temperature. On the one hand his value may be on the low side, as organic molecules and even life may survive at somewhat higher temperatures, as demonstrated e.g. by life forms found near thermal vents in the bottom of Earth's oceans. On the other hand, moist greenhouse runaway may start at average global surface temperatures as low as 355K (Wolf et al. 2017). At any case, our results are not very sensitive to minor variations of the temperature range.

The lower temperature limit is naturally chosen as the freezing point of water, which is only weakly dependent on pressure, and is widely accepted as a lower limit for organic processes. Of course also this limit is somewhat conservative, as high salinity can lower the freezing point. Still lower surface temperatures do not exclude liquid water and life under an ice cover, e.g. in the case of geothermal or tidal heat like in Europa and Encelladus. As mentioned above, tidal heating may be important in habitable planets of M-type stars (Dobbs, Heller and Turner 2017).

On synchronously orbiting planets the highest surface temperature, $T_0$, occurs at the sub-stellar point, $\theta=0$. When $b<1$ $T_0$ is given by eq. 6a. If $b>1$ it needs to be modified using eq. 11. The lowest surface temperature occurs at the far end of the night hemisphere, the opposite side of the substellar point, $\theta=180°$. When $b<1$ it is given by (eqs. 6a,b),

$$T_{min} = \left(\frac{f}{4-3f}\right)^{1/4} T_0 . \qquad (14)$$

When $b>1$ LHT cannot be neglected and $T_{min}$ must be calculated numerically. However, also for moderate values of $f$ and $b$ eq. 14 turns out to be a good approximation. Combining eqs. 6a,b, 12 and 14 the lowest and highest temperatures can be written as

$$T_{min} = 278\ (Hf)^{1/4}\ \text{K} \qquad (15a)$$

$$T_{max} = T_0 = 394\, H^{1/4}\, (1-\tfrac{3}{4}f)^{1/4}\ \text{K} \qquad (15b)$$

The corresponding values of the dimensionless temperature are $y_{max}=1$ and

$$y_{min} = \left(\frac{f}{4-3f}\right)^{1/4}. \qquad (16)$$

When the local heat transport is large ($b>10$) it dominates global heat redistribution (cf. Wordsworth 2015). In that case the planet is nearly isothermal: $y\sim 1$ and

$$T = 278\, H^{1/4}\ \text{K} \qquad (f\sim 1\ \text{or}\ b\gg 1) \qquad (17)$$

Figs. 7 and 8 show the highest and lowest surface temperatures as a function of the heat redistribution parameters $f$ and $b$, respectively, for a few values of the heating factor $H$. The $f$- and $b$-values of the terrestrial planets of the Solar System are taken from table 1.

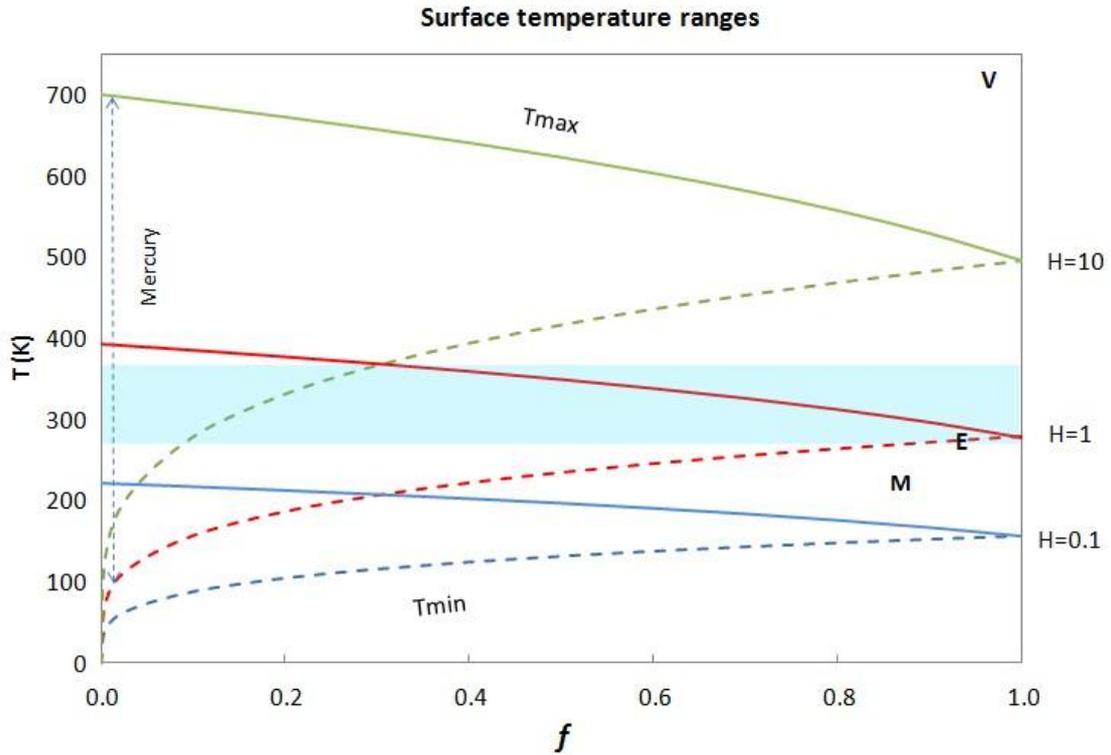

Fig 7. Highest and lowest surface temperatures $T_{min}$ (dashed) and $T_{max}$ (solid), vs. the global redistribution parameter, $f$ (with $b=0$), for three values of the heating factor $H$, marked on the right axis. The liquid water range for 1 bar is shown by the shaded blue area. Letters indicate the terrestrial planets of the Solar System.

For moderate LHT ($1<b<10$), the curves in fig. 8 were calculated numerically from eq. 10. Forhe assumption of $f=0$ in fig. 8 may not be realistic. When $b<1$ (low local heat redistribution), little energy reaches the night side, so its temperature would be near zero. This yields large temperature gradients which in turn induce global redistribution, producing an effective $f<0$. This effect is approximately described by an interpolation between the $f$ and *b* curves in the regime *1<b<10* (dotted curves in fig. 8). For weak local redistribution (*b<<1*) the dotted curves are set to the value given by eq. 15a replacing *f* by *b*.

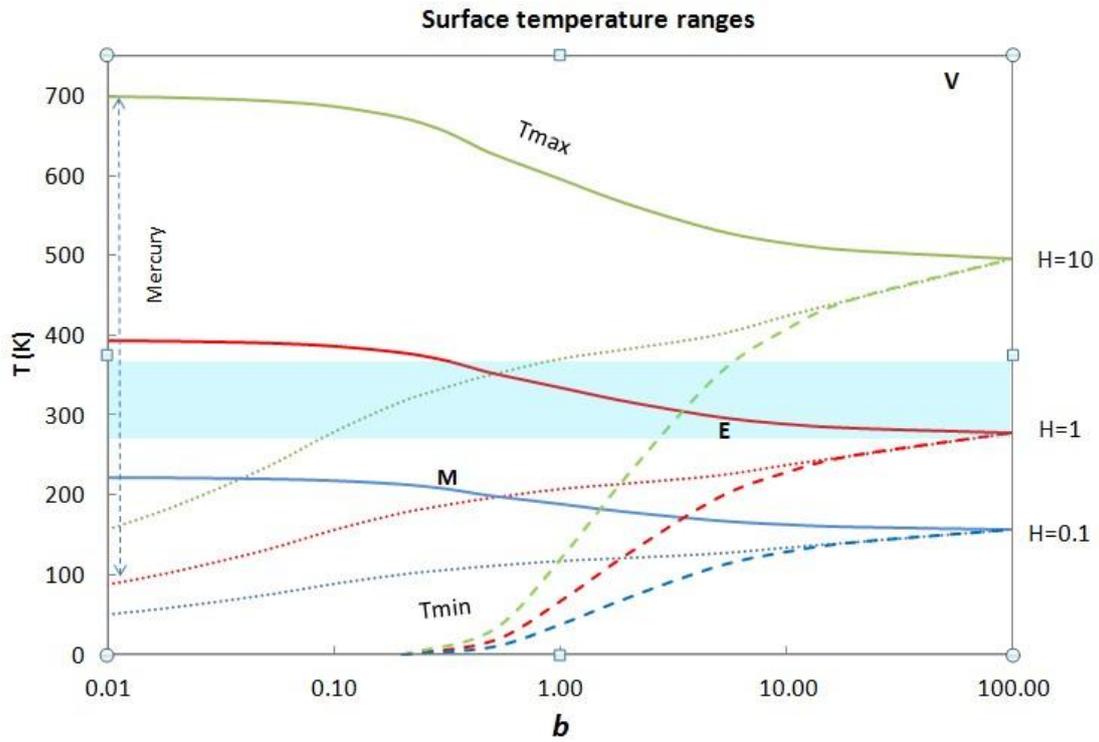

Fig. 8. Surface temperatures $T_{min}$ (dashed) and $T_{max}$ (solid), vs. the local transport parameter, *b*, for three values of the heating factor *H*, marked on the right axis. The dotted curves represent a more realistic approximation, allowing for a modest global redistribution when local heat transport is low (see text). The liquid water range for 1 bar is shown by the shaded blue area. Letters indicate the terrestrial planets of the Solar System.

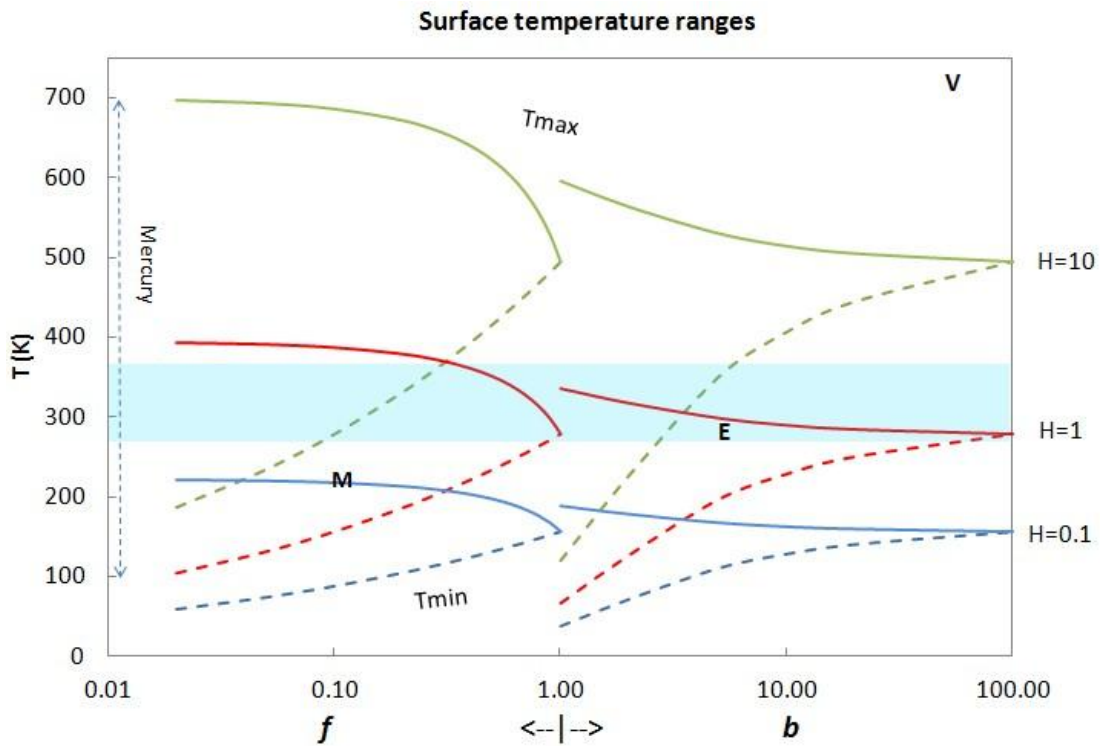

Fig. 9. A combined chart of the extreme temperatures on locked planets in the *H-f-b* parameter space. The temperatures $T_{min}$ (dashed) and $T_{max}$ (solid), vs. the global redistribution parameter *f* (left side, assuming *b=0*) and vs. the local transport parameter, *b* (right side, assuming *f=0*). The liquid water range for 1 bar is shown by the shaded blue area. Letters indicate the terrestrial planets of the Solar System, for Mars and Mercury *f* is assigned the value of *b*.

**7. The bio-habitable domain in the parameter space**

As described above, rather than using the spatial boundaries of the usual habitable zone, we define the range of surface temperatures allowing liquid water *and* the evolution of organic life, that is, complex organic molecules. The lower temperature boundary is obviously the freezing point of water, which only slightly depends on the pressure. The upper temperature boundary would be the highest temperature allowing complex organic molecules and life. On Earth extremophiles near hydrothermal vents have been observed to survive and multiply at temperatures as high as ~400K[6], but much higher temperatures would be iminical to complex organic molecules and processes. Depending on the atmospheric pressure, this may be lower or higher than the temperature allowing liquid water. Conservatively, we take as the temperature range allowing the evolution of life as 273K<*T*<373K. Minor variations of the upper boundary would slightly alter the range for the heating factor (eq. 18), but would not substantially change our results.

---

[6] https://www.ncbi.nlm.nih.gov/pubmed/9680332

Bio-habitability requires the coldest region on the planet's surface to remain below the upper limit, namely $T_{min}$ < 373K, and the hottest region to stay above freezing, that is, $T_{max}$ >273K. Note however that high salinity, water trapping and runaway warming or glaciation may modify these values, as discussed above. In terms of the heating factor $H$ and the global redistribution parameter $f$, this condition can be written as

$$\left[\frac{T_{\max}(f)}{273\text{K}}\right]^{-4} < H < \left[\frac{T_{\min}(f)}{373\text{K}}\right]^{-4}. \qquad (18)$$

Combining these two conditions with eqs. 14 and 15 gives

$$0.23(1-\frac{3}{4}f)^{-1} < H < 3.2f^{-1}. \qquad (19)$$

This range is shown in fig. 12 below. As $H=H_g s$ these relations may be transformed into a condition on the greenhouse heating (or equivalently on the optical depth $\tau_{LW}$) as a function of insolation or, for a given host luminosity, as a function of the distance from the host. With eq. 12 the boundaries in eq. 19 can be written in the form

$$0.23(1-\frac{3}{4}f)^{-1}a^2\left(\frac{L}{L_\odot}\right)^{-1} < H_g < 3.2f^{-1}a^2\left(\frac{L}{L_\odot}\right)^{-1}. \qquad (20)$$

The two frames of figure 10 show the bio-habitability ranges of the greenhouse factor for two values of host star luminosity. For clarity, instead of the greenhouse factor $H_g$ we plot $\tau_{LW}=H_g-1$ vs. the planet's distance from its host star. The allowed range corresponding to eq. 20 is the vertical span between the curves of highest possible greenhouse factor (red curves) and minimal value (blue) for a fixed value of the redistribution parameter $f$.

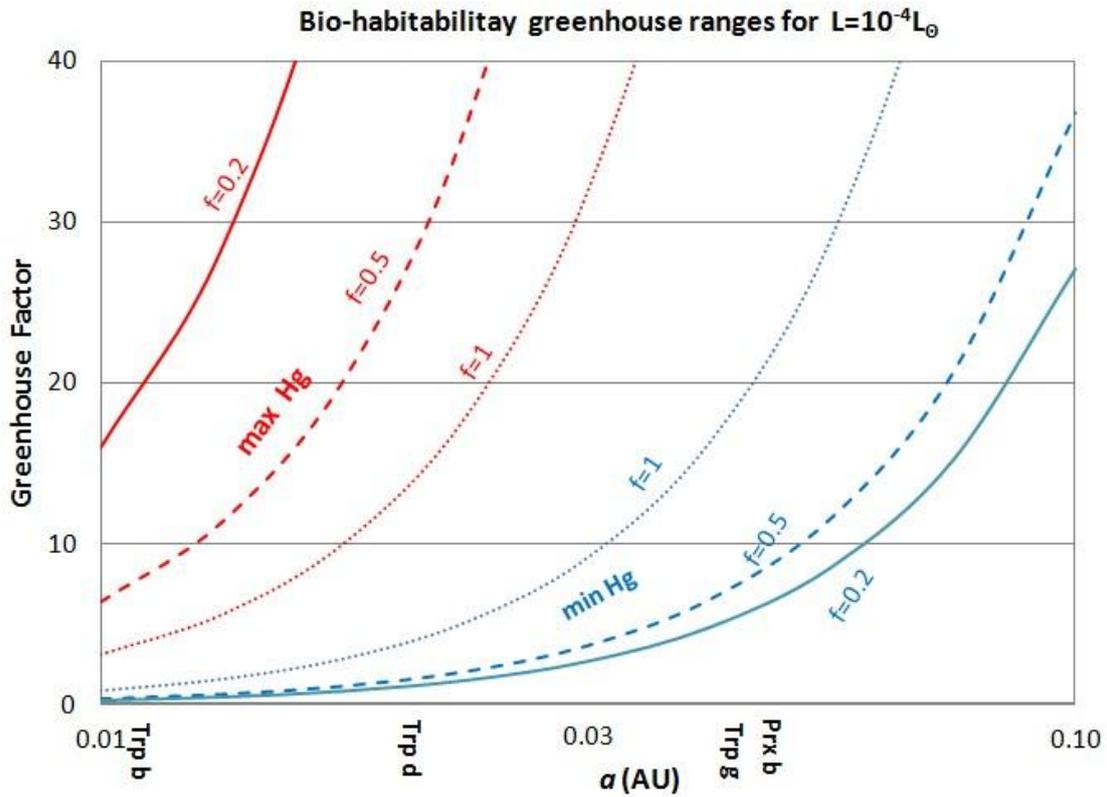

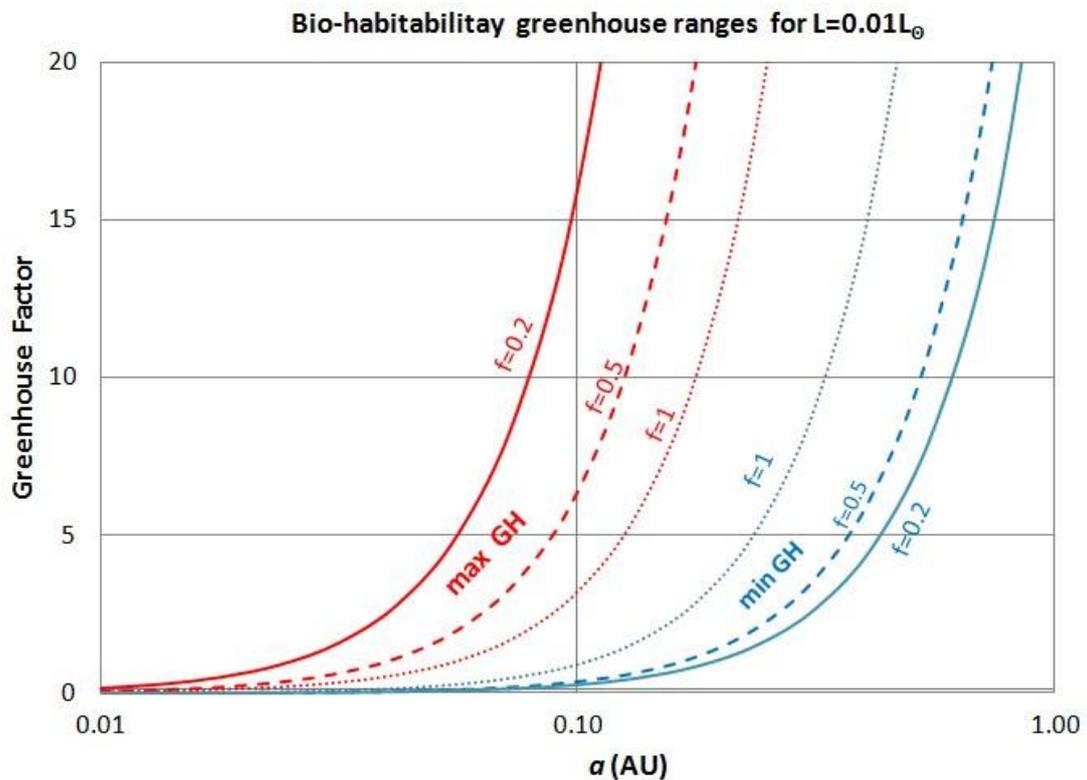

Fig. 10 Maximal (red) and minimal (blue) bio-habitable zone boundaries of the greenhouse factor ($\tau_{LW}=H_g-1$) vs. semi major axis, for three values of the heat redistribution parameter: $f$=0.2 (solid), 0.5 (dashed) and 1 (dotted). The host luminosity is assumed to be $L=10^{-4}L_\odot$ (upper frame) and $L=0.01L_\odot$ (lower frame). The locations of Proxima b and 3 of the Trappist-1 planets are marked on the x-axis of the upper frame.

The luminosity used in the upper frame of fig. 10 is of the order of that of Trappist-1[7], hence the orbital distances of the planets are indicated on the horizontal axis of that frame. For example, for Trappist-1d, the bio-habitable range with a redistribution of $f=0.5$ is $2<\tau_{LW}<30$, significantly larger than in the isothermal case ($f=1$), where $4<\tau_{LW}<15$.

Eq. 19 may be written in terms of the atmospheric heating and the insolation,

$$0.23(1-\frac{3}{4}f)^{-1}s^{-1} < H_{atm} < 3.2 f^{-1} s^{-1}. \qquad (21)$$

The corresponding ranges of the atmospheric heating factor can be seen in fig. 11, which shows the same boundaries as fig. 10 but in the parameter space defined by the atmospheric heating vs. the insolation (calculated with the bolometric luminosity). As in the previous example, the bio-habitable range for Trappist-1d with redistribution $f=0.5$ is $0.3<H_{atm}<6$, while the analogous range in the isothermal case is only $0.7< H_{atm} <2.5$.

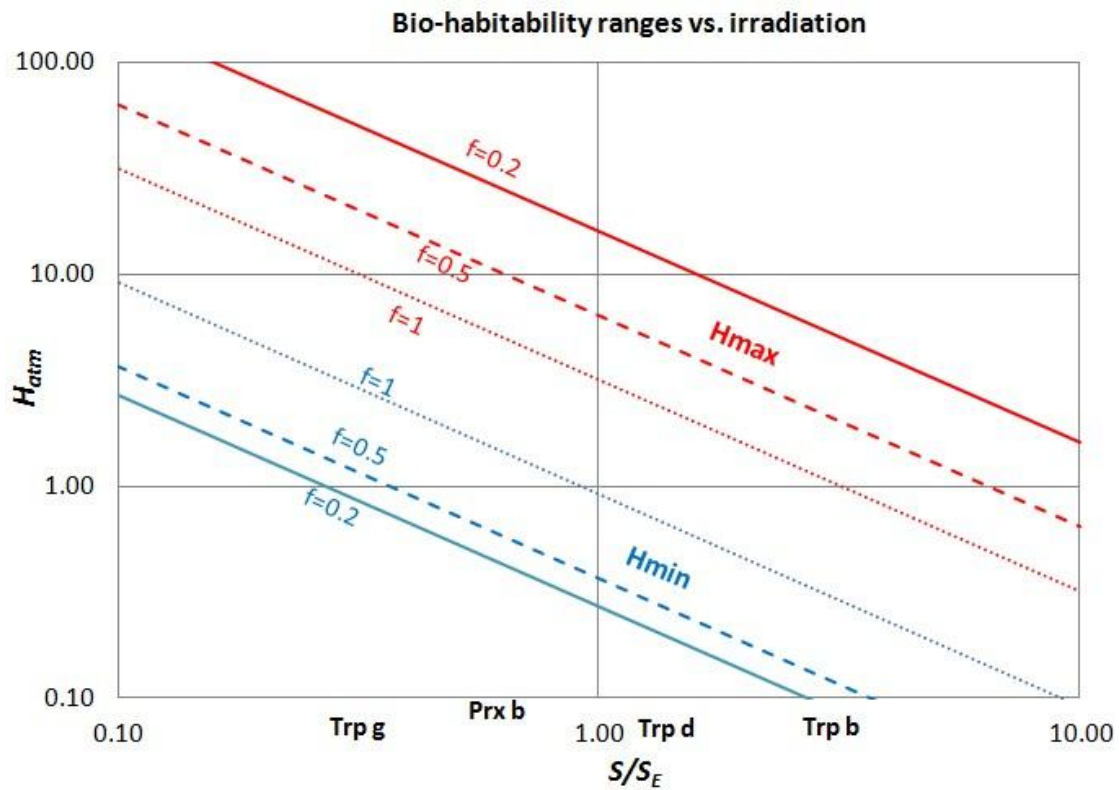

Fig. 11 Maximal (red) and minimal (blue) bio-habitable zone boundaries of the atmospheric heating factor vs. insolation, for three values of the heat redisribution parameter: $f=0.2$

---

[7] The bolometric luminosity of Trappist-1a is $0.0052 L_\odot$, but as most of it is in the long wavelength (visual luminosity only $3.7\ 10^{-6}\ L_\odot$ ), the radiation flux at the surface depends on $\tau_{LW}$ and may significantly lower; e.g. if $\tau_{LW}=3$ it would be $\sim 10^{-4}\ L_\odot$.

(solid), 0.5 (dashed) and 1 (dotted). The locations of Proxima b and 3 of the Trappist-1 planets are marked on the x-axis.

Fig 11 shows that in a significant part of the parameter space, spanned by the atmospheric properties (heating factor) and insolation, temperatures supporting liquid water and organic molecules may exist at least on part of the surface of locked planets. The size of this domain in the parameter space is smaller for nearly isothermal planets and significantly increases when global heat transport is less efficient.

Eqs. 18-19 also define the bio-habitable range of the heating factor, as a function of the global redistribution parameter, as shown in fig. 12. For example, for *f=0.2* eq. 19 gives *0.24<H<14*. This is consistent with fig. 4: on the curve *H=0.3* the highest temperature $T_{max}$ is barely above 273K, while for the curve *H=10* the lowest temperature $T_{min}$ is just below 373K.

While vertical lines in fig. 11 give the bio-habitable range of the atmospheric heating factor for a fixed insolation, in analogy to eq. 18 we may define the boundaries of the bio-habitable domain for varying insolations. Combining eq. 13 with the spatial boundaries of the habitable zone gives an insolation-independent condition in the parameter space defined by the heat redistribution parameters and the atmospheric heating factor. If we adopt the boundaries in the classical habitable zone, e.g. for the inner edge the insolation of Venus and the outer edge at the insolation of Mars, the insolation relative to Earth is in the range *0.52<s<2.3* (cf. the habitable zone boundaries and their dependence on the atmospheric pressure and composition, Forget 2013; Vladilo et al. 2015). Since $H_{atm} = H/s$ we have

$$H_{\min} s_{\max}^{-1} < H_{atm} < H_{\max} s_{\min}^{-1},$$

where $H_{min}$ and $H_{max}$ are the lowest and highest values of the heating factor (for a given value of *f* ) in eq. 19, that permit liquid water and organic molecules to exist at some point of the surface of a tidally locked planet. Substituting the Venus-to-Mars HZ-limits in eq. 19 gives

$$0.12 \left(1 - \frac{3}{4}f\right)^{-1} < H_{atm} < 7.4 f^{-1} \qquad (22)$$

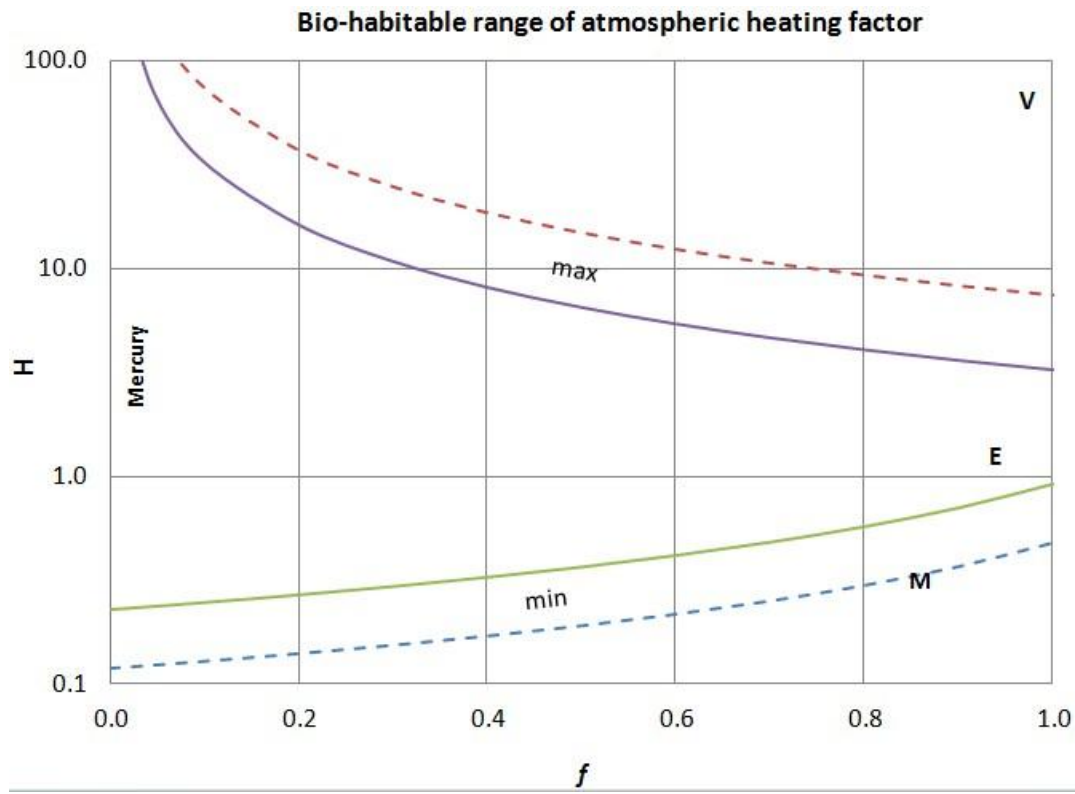

Fig. 12 Bio-habitable ranges of the heating factors vs. *f*. Solid curves denote *H*, dashed ones - $H_{atm}$. Upper curves indicate the maximal value, beyond which the planet is too hot, lower ones indicate the minimal value, below which the planet is too cold. Letters indicate the terrestrial planets in the Solar System.

A similar analyses can be done for the LHT parameter (*b*). When *1<b<10*, a condition on *H* analogous to eq. 19 may be calculated numerically using eq. 10, giving modified temperature ranges, as shown in fig. 13. For large LHT, b>>10, as well as for *f~1*, the planet is isothermal. In that case, eq. 17 can be substituted in eq. 18, yielding 0.93<*H* <3.2.

Similarly for the atmospheric heating factor, a condition analogous to eq. 21 may be calculated numerically when *1<b<10*, by using eq. 10. For *b>10* (as well as for *f ~1*) the planet is nearly isothermal and eq. 17 gives $0.40 < H_{atm} < 6.15$.

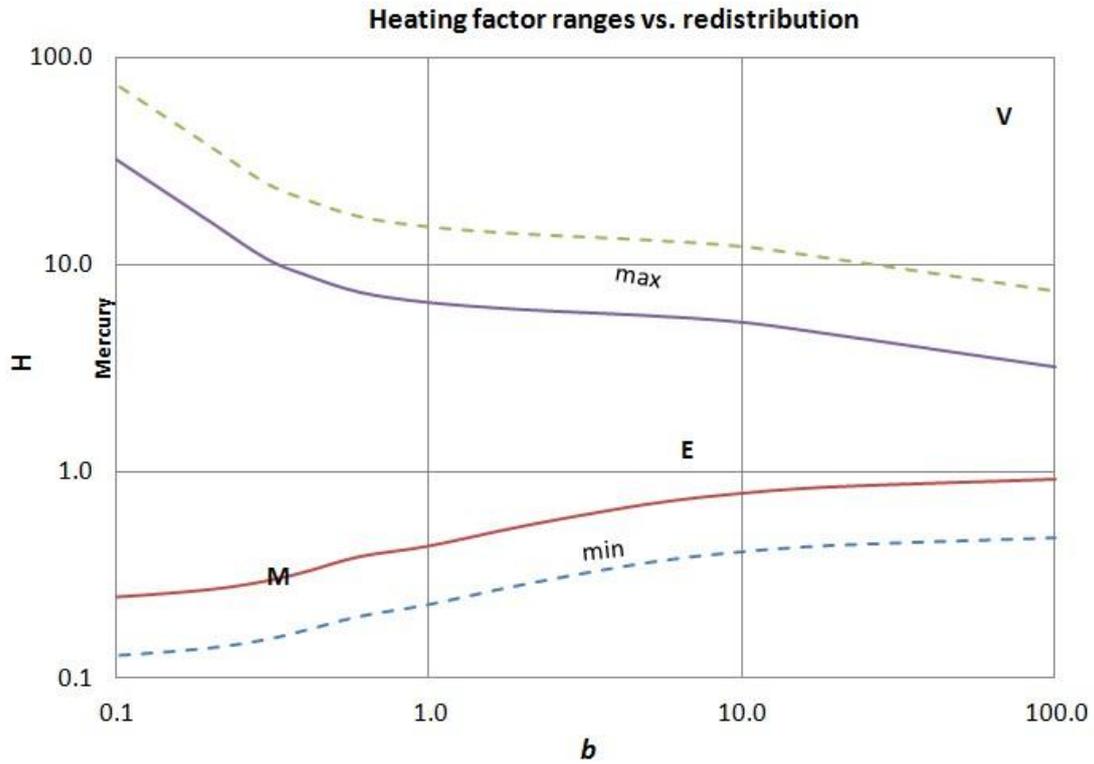

Fig. 13. Bio-habitable ranges of the heating factors vs. $b$. Solid curves denote $H$, dashed ones represent $H_{atm}$. The lower curves indicate the minimal value, below which the planet is too cold. The upper curves indicate the maximal value of $H$ or $H_{atm}$, beyond which the planet is too hot (see text). Letters indicate the terrestrial planets in the Solar System.

The upper curves (maximal values of the heating factors) in fig. 13 are calculated with the conservative night hemisphere temperatures (dotted curves in fig. 8).

The bio-habitable constraints on the heating factor may be used to constrain the allowed range of atmospheric heating factor for a specific planet. For example, consider Proxima b. Since Proxima b has $s = 0.65$, substitunig $H_{atm}=H/s$ in eq. 18 gives

$$0.35(1-\tfrac{3}{4}f)^{-1} < H_{atm} < 4.9\, f^{-1}. \qquad (23)$$

As in fig. 9, we note that a large fraction of the $H_{atm}$–$f$ or $H_{atm}$–$b$ parameter space apparently supports liquid water and bio-habitability on part of the planet's surface. This is true also for particular planets, as seen in fig. 14 for Proxima b.

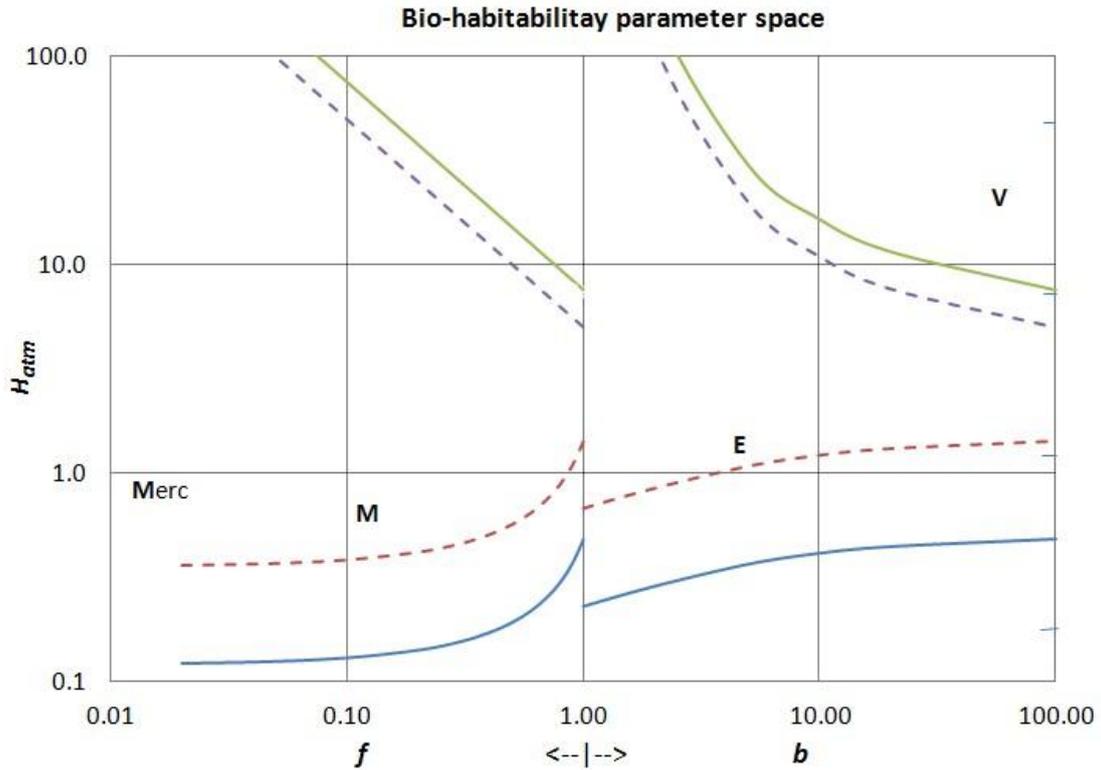

Fig 14. Combined chart of the conservative (273K<T<373K) life supporting temperature domain in the diagram of the atmospheric heating factor $H_{atm}$ vs. the heat redistribution parameters. Left panel: $H_{atm}$ vs. the global parameter $(0<f<1)$. Solid curves denote the boundaries of the bio-habitable domain (eq. 22). Right panel: $H_{atm}$ vs. the local transport parameter $(1<b<100)$ with $f=0$. The dashed curves refer to Proxima b (eq. 23). Letters indicate the terrestrial planets of the Solar System (for Mars and Mercury the $f$ coordinate is taken as the value of $b$).

## 8. Biosignatures of M-dwarf planets

The condition for life is often considered as liquid water on the planet surface. We have added the biotic temperature constraint, which may narrow the life-supporting region. In the previous section we have charted the parameter space of tidally locked M-dwarf planets, showing that such planets can have liquid water and life-supporting local surface temperatures for a wide range of atmospheric properties. On the other hand, as has been pointed out, liquid water may actually be absent even on planets within the habitable zone, due to processes such as water trapping, greenhouse runaway evaporation, snowball runaway, limit cycling or atmosphere erosion.

Liquid water may be a precursor of photosynthesis and atmospheric oxygen, also on planets of M-stars (e.g. Gale and Wandel, 2017). Spectral information about the planet's atmosphere may assess the presence of life, by identification of atmospheric biosignatures. While a large abundance of water vapor in the atmosphere of a planet may well indicate an ocean or liquid water in a limited part of the surface, an atmosphere oxygen is not necessarily a biotic indicator. Oxygen-rich atmospheres may evolve on M-dwarf planets by a-biotic processes giving a false positive oxygen signature (Domagal-Goldman et al. 2014; Harman et al. 2015, Luger and Barnes 2015). Oxygen may be produced by a-biotic processes such as photolysis of water or pre-main sequence evolution of the M-dwarf host (Tian 2009; 2014; Wordsworth and Pierrehumbert 2014). However, also massive photolysis has a spectral signature, as it would produce detectable byproducts (Meadows *et al.* 2016) such as $O_4$. To associate oxygen absorption features with photosynthesis it would be necessary to backup oxygenic signatures with other biomarkers (Seager *et al.* 2016) such as CHN and $CH_4$, or exclude the water photolysis mechanism by detecting significant quantities of non-condensing gases such as $N_2$. A relatively robust potential biosignature is considered to be the detection of oxygen ($O_2$) or ozone ($O_3$) simultaneous to methane ($CH_4$) at levels indicating fluxes from the planetary surface in excess of those that could be produced abiotically. For example, simultaneous detection of methane, ozone *and* $O_2$ may exclude abiotic production (Domagal-Goldman et al. 2014).

If the evolution of life on Earth is typical, a high level of oxygen in the atmosphere is reached 2-3 Gyr after the appearance of early life. As many red dwarfs are older than our sun, and as we have shown in the previous chapters, a wide range of atmospheres allow liquid water and organic life within the HZ of M-dwarfs, it is not unlikely that a large fraction of planets in the HZ of M-stars have oxygen-rich atmospheres of biotic origin.

Oxygen and other bio-signatures may be detected in transiting exoplanets by observing atmospheric absorption lines in the transmitted spectrum of the host star. The smaller the host star, the easier this can be accomplished, so M-dwarfs are the most suitable candidates. TESS is likely to find dozens of transiting Earth or Super earth-size planets possibly close enough for spectroscopic biosignature detection by JWST (Seager 2015).

By the same argument, it would be even easier to detect biomarkers in planets of white dwarfs (Loeb and Maoz 2013). Due to their low luminosity, habitable zones of white dwarfs are relatively small and hence also planets in the HZ of a white dwarf are likely to be tidally locked. It is likely that as much as 30% of the white dwarfs have planetary systems (Zuckerman et al 2010). However, because of the small dimensions of the host, transiting planets around white dwarfs may be less frequent, unless the planetary material is extended, as in the case of WD 1145+017 (Vanderburg et al 2015). Planets to near to a white dwarf would probably not be able to maintain their water and atmosphere over the violent past of their host, but planets may migrate from a larger distance or accrete new water from the debris of the parent planetary nebula faze.

### 9. The abundance of biotic planets

How abundant are habitable M-dwarf planets that actually have biotic oxygen? What is the fraction of planets with liquid water that develop oxygenic photosynthesis and detectable levels of oxygen? In order to tackle these questions we need first to estimate the abundance of candidates. Then, using the latest Kepler statistics, we estimate the relative abundance of planets for which we might find such biosignatures.

Assuming that biotic life is long lived, as on Earth (~4Gyr for mono-cellular life), the number of biotic planets can be expressed in a Drake-like equation (Wandel 2015),

$$N_b = N_* F_s F_{EHZ} F_b, \qquad (24)$$

where $N_*$ is the number of stars in the Galaxy, $F_s$ is the fraction of stars suitable for evolution of life and $F_{EHZ}$ is the fraction of such stars that have Earth-size planets within their habitable zone. The last parameter, $F_b$, is the (yet unknown) biotic probability, that a habitable planet actually becomes biotic within a few billion years. The average distance $d_b$ between biotic neighbor planets can then be shown to be (Wandel 2015; 2107)

$$d_b \sim 3 \, (n_* F_s F_{EHZ} F_b)^{-1/3} \text{ pc} \qquad (25)$$

where $n_*$ is the stellar number density in the solar neighborhood. For red dwarf stars, $n_* \sim 0.2$ pc$^{-3}$ and $F_s \sim 0.7$ (Winters et al. 2015, Henry et al. 2006).

The abundance factor $F_{EHZ}$ of Earth-size habitable planets can be estimated from the Kepler data (e. g. Dressing and Charbonneau 2015). Within a conservative definition of the habitable zone (based on the moist greenhouse inner limit and maximum greenhouse outer limit), there are $\{0.16\}_{-0.07}^{+0.17}$ Earth-size planets (1-1.5 $R_\oplus$) and $\{0.12\}_{-0.05}^{+0.10}$ super-Earths (1.5-2 $R_\oplus$) per red dwarf. With broader HZ boundaries (Venus–Mars insolation) these estimates increase by ~50%. This gives a minimum of 0.09 Earth-size planets to a maximum of 0.75 habitable Earth- or Superearth-size planets per red dwarf. Hence $F_{EHZ} \sim 0.1$-$0.75$, depending on the definition of the habitable zone.

Substituting these values in eq. 25 gives

$$d_b \sim (0.7\text{-}1.4) \, F_b^{-1/3} \text{ pc}. \qquad (26)$$

For example, assuming one in ten Earth size habitable planets of M-stars becomes biotic ($F_b = 0.1$), eq. 26 gives $d_b \sim$ 1.5-3 pc. The results of the previous sections could place the actual value of $F_{EHZ}$ closer to unity, as well as increase the expected value of $F_b$.

The expected number of biotic planets within a distance $d$ may be derived from eq. (25) by substituting the values of $n_*$ and $F_s$,

$$N_b(d) \sim 160 \, F_{EHZ} F_b \left(\frac{d}{10 \text{ pc}}\right)^3 \qquad (27)$$

Figure 15 shows the expected number $N_b$ as a function of the distance and the product $F_{EHZ} F_b$. The previous sections imply that the *effective* value of $F_{EHZ}$ (the fraction of Earth-sized planets in the bio-habitable zone) could be close to unity. Bio-habitability, as defined in sections 6-7 (surface temperatures supporting liquid water and organics), allows evolution of oxygenic photosynthesis[8] (Gale and Wandel 2017; Wandel and Gale 2017) and an oxygenic biosignature. This gives a direct relation between the observed number of planets with oxygenic signatures and the biotic probability $F_b$, with the caveat of non-biotic false-positive oxygen signatures discussed earlier.

TESS, scheduled for launch in 2018, is projected to find hundreds of transiting Earth-to-Superearth size planets, many of them in the habitable zones of nearby M-dwarfs (Ricker, *et al*. 2014; Seager 2015). By geometrical and statistical arguments, only a small fraction of the habitable red dwarf planets are expected to be transiting. The transiting angular range depends mainly on the size of the host star ($R_*$) and the distance of the planet from its host star ($a$). For $R_* \ll a$ the transiting probability is $\sim R_*/a$, which for planets in the habitable zone of a red dwarf turns out to be 1-2%. Hence we need to find a large enough number of transiting planets, in order to produce a statistically significant subset of candidates.

In order to assess the number of candidates, given the spectral signature detection distance appropriate for the particular method and platform, we can determine the number of expected candidates by calculating the cumulative number of biotic planets as a function of distance (Wandel 2017b).

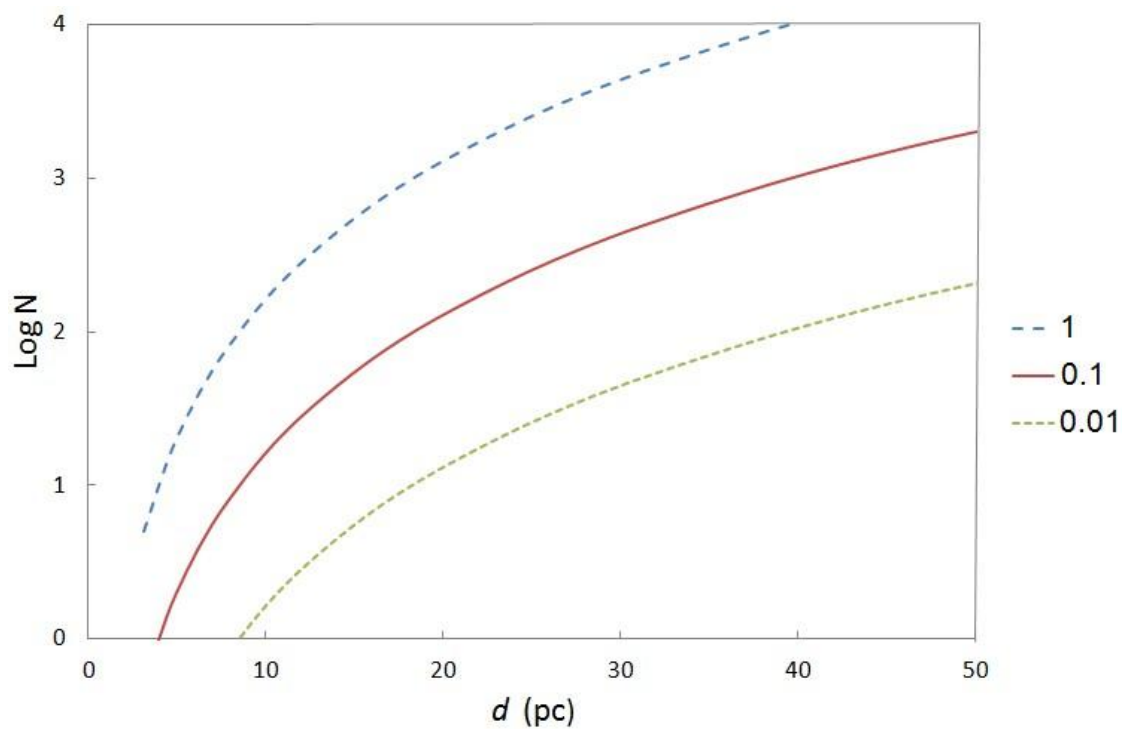

Figure 15. The number of red dwarf planets with oxygenic bio-signatures expected within a distance *d*, for several values of the product $F_{EHZ} F_b$ (marked on the right), where $F_b$ is the biotic probability *(see text) and* $F_{EHZ}$ is the fraction of M stars with Earth-sized habitable-zone planets.

---

[8] For gravitationally locked planets, photosynthesis may exclude the regime where only the night-hemisphere is bio-habitable, unless the atmosphere is sufficiently dense to scatter light from the day hemisphere.

From eq. 27 and in fig. 15 we can, substituting $F_b=1$, calculate the expected number of candidate planets within a given distance (the biosignature detection range) and subsequently estimate the number of transiting planets. If the M-dwarf habitability fraction is $F_{EHZ}=0.5$, one could expect ~2000 habitable planets of red dwarfs within 30pc. With a transiting probability of 1-2%, TESS can be expected to detect 20-40 transiting ones. As an example, let us assume that out of 30 candidates JWST finds oxygen signatures in 20 planets, and half of them have other supporting biosignatures. One could then estimate the biotic probability to be $F_b$~0.3±0.05. As JWST will probably be marginally capable of detecting biosignatures and may need long exposures, at least as a pilot it would be better to concentrate on a short list of best candidates (nearby systems with several planets in the bio-habitable zone, such as Trappist-1).

**Summary**

We use a simple climate model in order to investigate the surface temperature distribution and habitability of gravitationally locked planets, taking into account irradiation, albedo, horizontal heat transport and atmospheric effects such as screening and the greenhouse effect. We show that habitable-zone planets of M-dwarf stars may have temperatures supporting liquid water and complex organic molecules on at least part of their surface, for a wide range of atmospheric properties and heat transport. We apply these results to Proxima Cen b and to the Trappist-1 system and discuss the implications to searching oxygen and other biosignatures in transiting habitable planets of nearby M dwarfs (cf. Gardner et al, 2006; Seager 2015). From the Kepler data we estimate that within 30pc TESS may find ~10-40 transiting candidates. We show how detecting a few planets with atmospheric oxygen we may estimate the abundance of photosynthesis and biotic planets (or an upper limit, if none are detected) providing evidence for life (Spiegel and Turner, 2012).

Acknowledgements: the author thanks an anonymous referee for careful reviewing and useful comments, Joe Gale for useful discussions and the organizers of the Astrobiology 2017 and 51ESLAB meetings, where parts of this work have been presented.

# References

Alibert, Y., Benz, W. 2017, A&A 598, L5, e-print arXiv:1610.03460.

Anglada-Escudé, G., Amado, P. J., Barnes, J., *et al.* 2016, Nature, 536, 437.

Barnes, R, Deitrick, R., Luger, R. *et al.* 2016, e-print arXiv: 1608.06919v1.

N.M. Batalha, J.F. Rowe, S.T. Bryson, *et al.,* 2013, ApJ Suppl., 204, 24B, 24.

Carone, L., Keppens, R. and Decin, L. 2015, MNRAS 453(3): 2412-2437.

Davies, J. H., & Davies, D. R. 2010, Solid Earth, 1, 5

Dobbs, V., Heller, R. and Turner, E.L. 2017, A&A, e-print arXiv:1703.02447

Domagal-Goldman et al. 2014, ApJ., 807, 90.

Dressing C.D. and Charbonneau D. 2015, ApJ., 807, 45.

Froget, F. 2013, Int. J. of Astrobiology 12 (3): 177

Gale, J. and Wandel, A. 2017, Int. J. of Astrobiology, 16 (1), 1. eprint arXiv: 1510.03484.

Gardner J.P. *at al.* 2006, Space Sci Rev 123,485.

Gillon, M. et al. *Nature*. 542 (7642): 456. eprint arXiv:1703.01424.

Goldblatt, C. 2016, submitted to ApJ. e-print arXiv: 1608.07263.

Guillot, T. 2010, A&A 520, A27.

Haberle, R., McKay, C., Tyler, D., & Reynolds, R. 1996, in Circumstellar habitable zones., ed. L. Doyle, 29.

Haqq-Misra, J., Kopparapu, R.K., Batalha, N.E. et al. 2016, ApJ 827, 120.

Harman , C. E.; Schwieterman, E. W.; Schottelkotte, J. C.; Kasting, J. F.. 2015, ApJ 812, 127.

Heng, K., Kopparla, P. 2012, ApJ., 754, 60.

Heller R., Leconte J. & Barnes R. 2011, A&A, 528, A27.

Ingersoll A.P. 1969, J. Atmos. Sci. 26, 1191.

Joshi, M. M., Haberle, R. M., & Reynolds, R. T. 1997, Icarus, 129, 450

Kasting, J.F., Whitmire, D.P. & Reynolds, R.T. 1993. Icarus 101, 108.

Koll and Abbot 2016, ApJ 825, 99.

Komabayashi, M. 1967, *J. Meteor. Soc. Japan*. 45: 137.

Kopparapu, R. K. 2013, ApJ., 767, L8.

Kreidberg, L. and Loeb, A. 2016, ApJL, 832, 12.

Leconte, J., Forget, F.; Charnay, B.; Wordsworth, R.; Selsis, F.; Millour, E.; Spiga, A. 2013, A&A 554, 69L.

Leconte, J., Wu, H., Menou, K., & Murray, N. 2015, Science, 347, 632L.

Loeb, A. and Maoz, D. 2013, MNRAS., 432, pp. L11.

Luger, R., and Barnes*,* R. 2015. Astrobiology 15, 119.

Meadows, V.S., Arney, G,N. Schwieterman, E.W., et al. 2016, Submitted to Astrobiology, eprint, arXiv:1608.08620

Menou, K. 2013, ApJ 774, 51.

Merlis, T.M. & Schneider, T. 2010, J. Adv. Model. Earth Syst., 2010, 2, Art. #13, e-print arXiv:1001.5117

Mills, S. M., & Abbot, D. S. 2013, ApJL, 774, L17.

Owen ,J.E. & Mohanty, S. 2016, MNRAS, 459, 4088.

Pierrehumbert, R. 2011a, ApJ 726, L8.

Pierrehumbert, R. T. 2011b, Principles of Planetary Climate (Cambridge: Cambridge Univ. Press).

Ribas, I., Bolmont, I., Franck, S., et al. 2016, A&A, eprint arXiv: 1608.06813.

Ricker, G.R., Winn, N.J., Vanderspek, R. *et al*. , e-print arXiv:1406.0151.

Scalo, J., Kaltenegger, L., Segura, A. et al. 2007, Astrobiology, 7(1): 85.

Seager, S., and Deming, D. 2010, ARA&A, 48, 631.

Seager, S. 2015, PNAS 111, 35, 12634–12640, doi: 10.1073/pnas.1304213111.

Seager S., Bains W. and Petkowski J.J. 2016, Astrobiology 16(6), 465.

Selsis, F., Wordsworth, R. D., & Forget, F. 2011, A&A, 532, A1.

Snellen, I., de Kok, R.; Birkby, J. L.; Brandl, B., *et al.* 2015, A&A, 576, A59.

Spiegel, D.S. and Turner, E.L. 2012 PNAS 109, no. 2, 395.

Tarter, J.C. Backus, P.R. Mancinelli, R.L., *et al.* 2007, Astrobiology, 7: 30.

Tian, F. 2009, ApJ, 703, 905.

Tian, F., France, K., Linsky, J.L., Mauas, P. J. D., & Vieytes,M.C. 2014, Earth and Planetary Science Letters, 385, 22-27.

Turbet, M. Leconte , J., Selsis F., *et al.* 2016, A&A 596, 112. e-print arXiv: 1608.06827.

Vanderburg et al, 2015, Nature 526, 546.

Vladilo, G., Silva, L., Murante, G. et al. 2015, ApJ 804, 50.

Wandel, A. 2015, Int. J. of Astrobiology, 14, 511.

Wandel, A. 2016, in "Search for Life: from early Earth to Exoplanets", XII rencontres du Vietnam, https://www.youtube.com/watch?v=gJaz6jim4vs

Wandel, A. 2017a, Acta Astronautica 137, 498-503, e-print arXiv:1612.03844.

Wandel, A. 2017b, in 51st ESLAB symp. on "extreme habitable wolds", http://old.esaconferencebureau.com/docs/default-source/17a13-docs/eslab-2017-abstract-book-version-13122017.pdf?sfvrsn=0, p. 157.

Wnadel, A. and Gale, J. 2017 in "Astrobiology 2017", https://www.youtube.com/watch?v=BtDv-o0b_0c

Wang Y., Tian, F. and Hu, Y. 2014, ApJ 791, L12.

Wolf, E. T. 2017, ApJ 839, L1.

Wolf, E. T., Shields, A. L., Kopparapu, R. K. et al. 2017, ApJ 837, 107.

Wordsworth, R. and Pierrehumbert R. 2014, ApJ. 785, L20.

Wordsworth, R. 2015, ApJ 806, 180.

Yang, Y., Yonggang, L., Yongsun, H. and Abbot, D.S. 2014, ApJ. 796, L22.

Yang, Y., Leconte, J., Wolf, E. et al. 2016, ApJ. 826, 222.

Zuckerman, B., Melis, C.; Klein, B.; Koester, D.; Jura, M. 2010, ApJ 722, 725.